\documentclass[pre,twocolumn,superscriptaddress]{revtex4-1}

\usepackage[toc,page]{appendix}
\usepackage{color}
\usepackage{enumerate}
\usepackage{amsfonts}
\usepackage{amssymb}
\usepackage{mathrsfs}
\usepackage{amsmath}
\usepackage[mathscr]{eucal}
\usepackage[english]{babel}
\usepackage{graphicx}
\usepackage{multirow}

\newcommand{\var}{\mathrm{var}}

\newcommand{\Vig}{{V_\mathrm{I}}}
\newcommand{\pig}{{\phi_\mathrm{I}}}

\newcommand{\Sshell}{S_\mathrm{shell}}

\newcommand{\rhofs}{\rho_f^*}
\newcommand{\e}{\epsilon}
\newcommand{\sweep}{c}
\newcommand{\matho}{\mathscr{O}}

\newcommand{\phid}{\phi_\mathrm{d}}

\newcommand{\phik}{\phi_\mathrm{K}}
\newcommand{\p}{\phi}
\newcommand{\prcp}{\p_\mathrm{rcp}}
\newcommand{\phij}{\phi_\mathscr{J}}
\newcommand{\Sc}{S_\mathrm{c}}

\newcommand{\Stot}{{S}}
\newcommand{\Svib}{S_\mathrm{vib}}

\newcommand{\Srti}{S_{\scalebox{.6}{\rm RTI}}}
\newcommand{\kb}{k_B}
\newcommand{\rhof}{\rho_f}

\newcommand{\Ng}{{\mathcal{N}_g}}
\newcommand{\Rs}{R_{\mathrm{shell}}}

\begin{document}
\title{Measuring glass entropies with population annealing}
\author{Christopher Amey}
\affiliation{Department of Physics, University of Massachusetts,
Amherst, Massachusetts 01003 USA}
\author{Jonathan Machta}
\affiliation{Department of Physics, University of Massachusetts,
Amherst, Massachusetts 01003 USA}
\affiliation{Santa Fe Institute, 1399 Hyde Park Road, 
Santa Fe, New Mexico 87501, USA}

\begin{abstract}
    We combine population annealing Monte Carlo and several thermodynamic integration techniques to measure equilibrium vibrational and configurational entropies in the metastable fluid regime beyond the dynamic glass transition.  We obtain results for a three-dimensional binary mixture hard sphere system. Our results suggest that the configurational entropy vanishes before the equilibrium pressure diverges, which implies that an underlying thermodynamic glass transition exists for this system.  The computational methods are general and can be applied to a variety of glass forming systems but are restricted to small system sizes. 
\end{abstract}

\maketitle

\section{Introduction}
    Configurational glasses have been an active area of research for decades, however, open questions about the nature of the glass transition remain. Broadly speaking, a glass transition corresponds to a dynamical phenomenon where, upon cooling, a fluid suddenly exhibits extremely slow dynamics and effectively behaves as an amorphous solid. The glass transition is, by convention, defined as the temperature where the viscosity reaches $10^{12}$ Poise~\cite{berthier_revmodphys}. This temperature does not correspond to a thermodynamic transition and was chosen from a practical standpoint: fluids with viscosities much larger than $10^{12}$ Poise can not be equilibrated within reasonable lab timescales. While the conventionally defined glass transition is somewhat arbitrarily defined, there are reasons to believe that the onset of slow dynamics may be a precursor to a true thermodynamic glass transition at some lower temperature. The nature of such a transition and whether it even exists is a central question in glass physics~\cite{kauzmann:48, gibbs:57, adam_gibbs:65, goldstein:69, kirkpatrick:87, kirkpatrick:87b, kirkpatrick:87c}.
    
    In this paper we apply microcanonical (NVT) population annealing Monte Carlo and thermodynamic integration to a binary hard sphere fluid in the glassy regime beyond the dynamic glass transition.
    Binary hard sphere fluids have been studied numerically~\cite{speedy:98,Angelani:2007, OdBe11, berthier_witten:09,Brambilla09,berthier:2010} and experimentally~\cite{Brambilla09, masri:09} in the past but, despite the relatively large body of research, a detailed understanding of the physics beyond the dynamic glass transition remains elusive. The hard sphere potential is well-suited to theoretical and numerical studies due to its simple functional form and a binary mixture can be designed to prevent homogeneous crystallization so that it displays a robust glassy regime. In hard sphere systems, temperature is not an independent control parameter and instead the inverse dimensionless pressure, $1/Z=N\kb T/PV$, is the relevant thermodynamic variable~\cite{berthier_2019}. In most simulations, the dimensionless pressure is an observable and the density or packing fraction, $\p$, is the control parameter. Therefore, the picture one should have in mind for hard sphere glasses is that as the density is increased, the dimensionless pressure increases and eventually diverges once the configuration of particles can no longer be compressed at the random close packing density, $\prcp$~\cite{liu:2007}.  
    
    There are many different theories that attempt to describe the physics behind glassy fluids at high density. One explanation, originally proposed by Gibbs and DiMarzio~\cite{gibbs_56, gibbs:57}, is that there exists a dynamic ergodicity breaking transition where phase space separates into many dynamically distinct regions at some density, $\phid$. In this picture, for $\phi < \phid$, the system behaves as a fluid that will ergodically sample all of phase space. After the dynamic transition, the mixing is no longer sufficient to equilibrate and phase space breaks into different regions, each of which corresponds to a different glass state. In terms of actual degrees of freedom, particles are unable to diffuse on large length-scales and, instead, remain trapped locally in cages composed of their neighbors in which they vibrate. In this picture, the phase space volume of each glass state corresponds to the local vibrational degrees of freedom of the particles and, furthermore, each glass state has a corresponding ``glass'' or ``vibrational'' entropy resulting from these local particle vibrations. In this paper we use the term ``glass state'' to refer to a localized region in phase space and ``particle configuration'' or ``configuration'' to refer to a specific set of particle positions within a thermodynamic state. In this terminology, a glass state is made up of many different configurations which are accessible to one another on relatively short time scales. The physics of such a system is dictated by a  ``configurational entropy per particle'', defined as 
    \begin{align}
        \label{eq:Sc}
        \Sc = \frac{1}{N} \log \Ng,
    \end{align}
    where $\Ng$ is the number of glass states.
    Understanding how $\Sc$ decreases as the density is increased and, ultimately, when $\Sc$ goes to zero is paramount to understanding whether or not a thermodynamic glass transition exists. If $\Sc$ goes to zero at a ``Kauzmann'' density, $\phik < \prcp$, then this would signal that a thermodynamic transition exists. Therefore, measuring $\Sc$ and extrapolating it to zero is a fundamental way to determine the high-density equilibrium behavior of supercooled fluids. In the Adam-Gibbs picture~\cite{adam_gibbs:65}, the Kauzmann transition results in the structural relaxation time increasing exponentially with $1/(T\Sc)$. As such, another tactic to estimating the value of $\phik$ is to make fits to dynamical quantities, such as structural relaxation times or viscosities, as a function of packing fraction in order to estimate the location of a dynamic divergence corresponding to the thermodynamic transition.
    
    Another possible scenario is that there is no thermodynamic transition at finite $Z$ and that $\phik=\prcp$. In this picture, the configurational entropy vanishes at $\prcp$ and, as a result, no thermodynamic transition would occur~\cite{OdBe11, berthier_2019}. In a thermal glassy system, this would be analogous to a super-cooled fluid branch ending at $T=0$ without a thermodynamic glass transition occurring at some $T_\mathrm{K} > 0$. 
    
    Much work has been done to determine the nature of the glass transition and to estimate the values of $\phik$ and $\prcp$ in hard sphere systems. Past work includes fits to dynamical functions, such as the Vogel-Fulcher-Tammann law, \cite{berthier_witten:09, Brambilla09, masri:09}, free volume fits \cite{OdBe11,berthier:2010, Callaham}, and direct measurements of the configurational entropy \cite{speedy:98, Angelani:2007}. 
    The primary obstacle in these works is in sampling equilibrium glass states beyond the dynamic transition, where known numerical methods are not efficient for binary systems and, as of yet, there is no consensus on whether a thermodynamic glass transition exists for three-dimensional binary fluids. Recent work with continuously polydisperse hard sphere systems have been able to probe unprecedentedly high densities while remaining in statistical equilibrium~\cite{Berthier:2016,Berthier:2017,Berthier:2018,berthier_2019,ozawa:2019,baranau_tallarek:2020}.  These studies suggest that a thermodynamic transition does exist in three dimensions. However, it is not clear that the physics of continuously polydisperse systems is the same as that of binary mixtures and, additionally, there are non-trivial issues with taking the thermodynamic limit of continuously polydisperse systems that need careful consideration. 
    
    The paper is organized as follows.  We begin by briefly reviewing the binary hard sphere model and the observables of interest in Sec.\ \ref{sec:model}. 
    We then describe the NVT version of the population annealing algorithm and introduce two new thermodynamic integration techniques to calculate the vibrational entropy of a glass state in Sec.\ \ref{sec:methods}.
    We present the results from large-scale simulations in Sec.\ \ref{sec:results} and the paper closes with a discussion in Sec.\ \ref{sec:disc}. 
    
\section{Model and observables} \label{sec:model}
    The system we study is a binary hard sphere fluid with a 50:50 mixture of particles with radius ratio 1.4:1~\cite{torquato_2012,OdBe11, Callaham}. This system is known to be a good glass former because although its high density equilibrium state in the thermodynamic limit is two monodisperse crystals separated by a domain wall, this state is inaccessible in simulations starting from a random mixture. One of the primary observables of interest in a simple fluid is the dimensionless pressure, $Z$, defined as
    \begin{align}
        \label{eq:EOS}
        Z = \frac{P V}{N \kb T},
    \end{align}
    where $P$ is the pressure, $V$ is the volume, $N$ is the total number of particles, $\kb$ is Boltzmann's constant, and $T$ is the temperature. It is common to use the packing fraction, $\phi$, as a control parameter where
    \begin{align}
        \phi =  N \frac{4 \pi r^3}{3V},
    \end{align}
    and $r^3 = (r_0^3 + r_1^3)/2$ is the average of the cubed radii of the two species. Because there is no potential energy in this system, the temperature only sets the average kinetic energy of the particles and the remaining physics depends on the dimensionless ratio $Z$. Thus, without loss of generality, we set $\kb = T = 1$. We work in the NVT ensemble where $N$ and $\p$ are set and $Z(\p)$ is measured. Although free energy is the thermodynamic potential for NVT, due to the triviality of the energy, all of the equilibrium physics is contained in the entropy as a function of $N$ and $\p$. 
    
    The binary fluid equation of state is well-approximated throughout the fluid phase by the phenomenological Boubl\'{i}k-Mansoori-Carnahan-Starling-Leland (BMCSL) equation of state~\cite{BMCSL_2, BMCSL_1}
    \begin{align}
        \label{eq:BMCSL}
        Z_\mathrm{BMCSL} = \frac{(1+\phi+\phi^2)-3\phi(y_1+y_2\phi)-y_3\phi^3}{(1-\phi)^3},
    \end{align}
    where $y_i$ are constants that depend on the polydispersity. For a 50:50 mixture of 1.4:1 size particles, $y_1 = 0.0513$, $y_2=0.0237$, and $y_3=0.9251$. This equation of state is very accurate when compared to numerical data~\cite{Callaham}, but it must break down at high density, as is clear since it remains finite until $\phi=1$.
    
    For glassy systems, we are ultimately interested in the configurational entropy, $\Sc$. There are several ways of estimating $\Sc$ directly, including counting inherent structures~\cite{goldstein:69} and using the Franz-Parisi potential~\cite{franz-parisi}.  However, the most common procedure, used here, is to measure the total entropy and vibrational entropy and then use the relation, 
    \begin{align}
        \label{eq:entropy_relation}
        \Stot = \Svib + \Sc,
    \end{align}
    where $\Stot$ and $\Svib$ are the total and vibrational entropies, respectively. 
    
    To better understand the definitions and relationships of these entropies, we start with the assumption that in the glassy regime configuration space is broken into ergodically disconnected regions, each of which corresponds to a different glass state. In the NVT ensemble each glass state $\nu$ appears with probability,
    \begin{align}
        w_\nu(\phi) =  \frac{\Omega_\nu}{\sum_{\tilde{\nu} \in c(\phi)} \Omega_{\tilde{\nu}}},
    \end{align}
    where the sum is over the set of all glass states, $c(\phi)$, at packing fraction $\phi$, and $\Omega_\nu$ is the configuration space volume of glass state $\nu$. The entropy of glass state $\nu$ is given by the standard definition~\cite{berthier_revmodphys},
    \begin{align}
        S_\nu(\phi) = \frac{1}{N} \log \Omega_\nu.
    \end{align}
    Throughout this work all entropies are  defined per particle.
    We define $\Svib$ as the average over the   entropies of the glass states,
    \begin{align}
        \label{eq:svib}
        \Svib(\phi) = \sum_{\nu \in c(\phi)} w_\nu(\phi) S_\nu(\phi).
    \end{align}
    The total entropy is given by the standard definition,
    \begin{align}
        \Omega(\phi) &= \sum_{\nu\in c(\phi)} \Omega_\nu(\phi), \\
        \Stot(\phi) &=\frac{1}{N} \log \Omega(\phi).
    \end{align}
    Using these definitions and Eq.\ \ref{eq:entropy_relation} yields an equation for the configurational entropy,
    \begin{align}
        \label{eq:Sc_theory}
        \Sc = -\frac{1}{N}\sum_{\nu\in c(\phi)} w_\nu(\phi) \log w_\nu(\phi),
    \end{align}
    which is similar in form to the Shannon entropy. If we assume that there is a finite number of glass states, $\Ng$, each of which has the same statistical weight, $w_\nu = 1/\Ng$, then $\Sc$ reduces to the standard definition given in Eq.\ \ref{eq:Sc},
    \begin{align}
        \Sc = \frac{1}{N} \log \Ng.
    \end{align}
    
    The pressure and the total entropy obey the standard thermodynamic relation which, in terms of packing fraction, is given by
    \begin{align}
       Z = -\phi\,\frac{\partial \Stot}{\partial \phi}.
    \end{align}
    Using this relation, the dimensionless pressure can be integrated with respect to the packing fraction in order to obtain the entropy, see App.\ \ref{app:entropy_norm} for details. Numerically measuring the vibrational entropy can also be done using thermodynamic integration but is non-trivial and is  discussed below in Sec.\ \ref{sec:vibrational}.  

\section{Computational methods}  
    \label{sec:methods}
    \subsection{Microcanonical population annealing} \label{sec:fl_PA}
        We simulate the binary hard sphere mixture at high density using an NVT ensemble version of population annealing Monte Carlo, first described in Ref.~\cite{Callaham}. Population annealing (PA) is a sequential Monte Carlo method~\cite{HuIb03, machta:10, Wang2015, amey2018,rose2019, weigel2021understanding} similar to nested sampling~\cite{nested_sampling:06,nested_sampling:14_01,nested_sampling:14_02} that is used to simulate systems with rough free energy landscapes. The general idea of population annealing is to initialize a large ensemble of independent simulations in an easy-to-equilibrate region of parameter space and to anneal towards a difficult-to-equilibrate region. After each annealing step, the population is resampled so as to keep the distribution in equilibrium. Initially the population can be kept in equilibrium via conventional Monte Carlo schemes, however, eventually the simulation enters a regime where the system is unable to equilibrate dynamically and resampling becomes necessary to keep the population in equilibrium.
        
        In this work, we deal with hard spheres in the NVT ensemble. All allowed configurations of hard spheres have the same energy and the roughness of the free energy landscape in the glassy regime is entirely due to a rough entropy landscape. In the NVT version of PA, each replica in the population is independently initialized as a gas of particles at low packing fraction. An equilibrating procedure such as molecular dynamics or Markov chain Monte Carlo is then applied to each member of the population. Here we use event chain Monte Carlo (ECMC)~\cite{Krauth:2009,Michel2014}. This step is performed to equilibrate and decorrelate the population, however, in the glass regime, it only manages to move particles within local cages. After running Monte Carlo, the population is annealed by increasing the packing fraction, $\phi$, following an annealing schedule, $\{ \p_0, \p_1, \ldots, \p_f \}$, where $\p_0$ is in the low density fluid phase and $\p_f$ is the highest packing fraction simulated in the glassy regime. The physical process of annealing corresponds to decreasing the box volume but, in simulations, it is computationally simpler to increase the sphere radii while keeping the volume fixed. As we will see below, these two annealing processes result in different changes in entropy so a correction must be made when performing thermodynamic integration.
        
        After increasing the sphere radii, a fraction of the population's configurations have overlaps and are illegal at the new density. Replicas with illegal configurations are erased or ``culled'' and are replaced by randomly resampling the remaining legal replicas with equal weight. The fraction of the population that is culled, $\epsilon$, is called the ``culling fraction'' and is an important quantity for setting the annealing schedule and integrating the entropy. After culling and resampling, this process is then repeated at the new packing fraction and annealing continues until $\p_f$ is reached.
        
        \begin{figure*}
            \centering
            \includegraphics[width=0.7\textwidth]{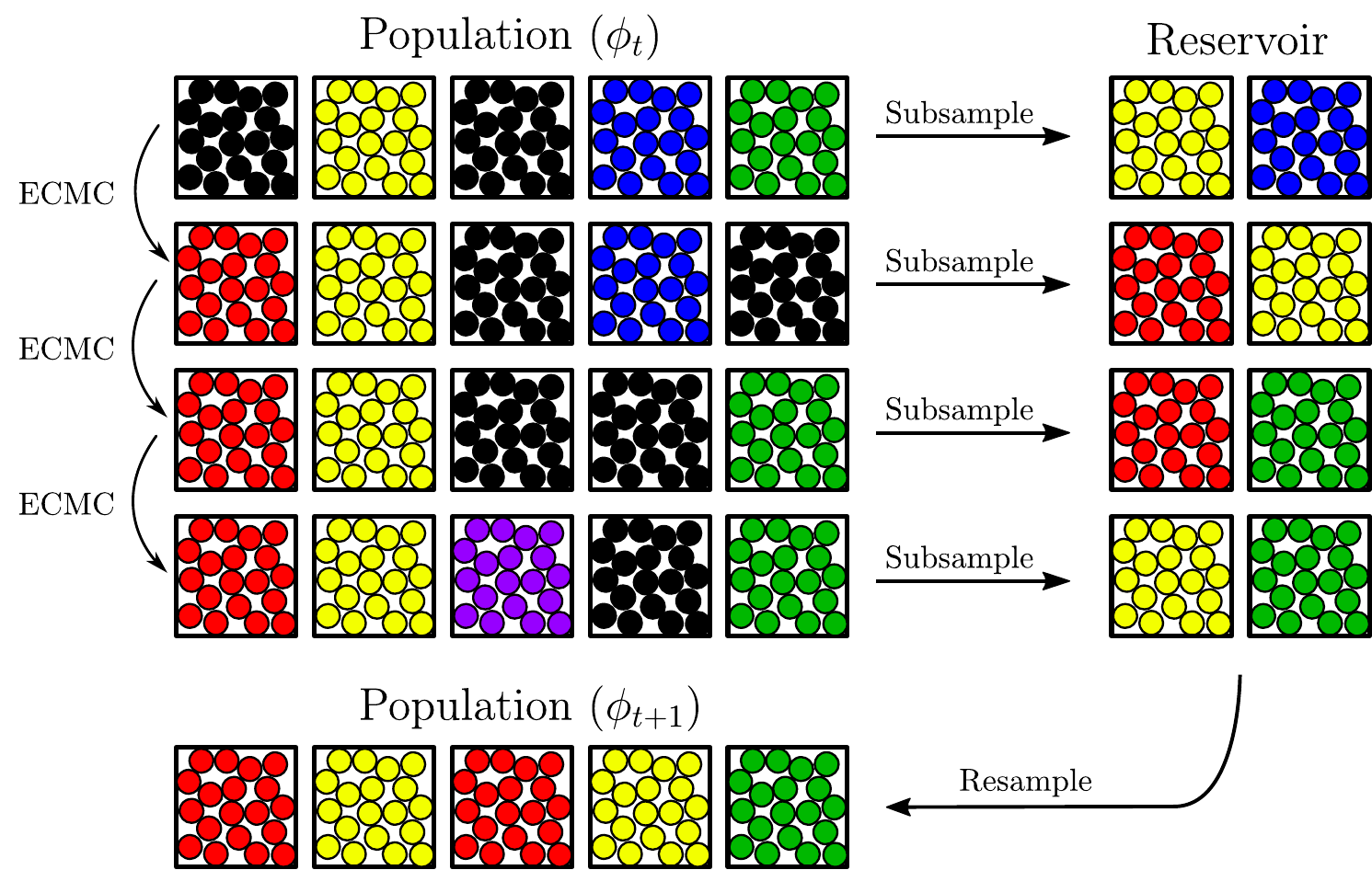}
            \caption{Diagram representing one hybrid microcanonical population annealing step. The packing fraction is fixed at $\phi_t$ and the population is equilibrated using ECMC for several intervals. After each interval, a fraction of the population is subsampled and configurations that are legal at the next packing fraction in the annealing schedule, $\phi_{t+1}$, are saved in a reservoir. Illegal configurations (marked black) are not saved. After equilibration is completed, the reservoir is randomly sampled to produce a new population at packing fraction $\phi_{t+1}$.
            }
            \label{fig:hybrid}
        \end{figure*}

        In this work we use a ``hybrid'' resampling method that is similar to that in Ref.~\cite{rose2019} in order to reduce the statistical errors associated with the resampling process. In the PA scheme of Ref.\ \cite{Callaham}, described above, the new population is resampled from the final population at the end of the annealing step. 
        The hybrid method decreases the statistical errors associated with resampling by increasing the frequency of sampling during a single annealing step. Instead of resampling the population once at the end of the equilibration process, the population is sampled several times and additional sweeps are performed before each sampling step. As a result, the entropy of each member of the population is taken into account several different times. This procedure is shown in Fig.\ \ref{fig:hybrid}. The scheme used in this work is to perform a total of $\sweep_k$ sweeps of ECMC on each member of the population at annealing step $k$, where we define the number of sweeps as the number of particle movements divided by the number of particles. The sweeps are broken into an initial burn in of $\sweep_k /2$ sweeps. After each subsequent sweep, a fraction of the population is subsampled and configurations that are legal at the next packing fraction are saved into a reservoir. In this work, the subsampled population consists of $R /(\sweep_k/2)*1.5$ randomly chosen replicas. Finally, at the end of the annealing schedule, $R$ replicas are chosen at random from the reservoir to represent the population at the beginning of the next annealing step. 
        
        The subsample size was chosen with consideration of the culling fraction in order to ensure that the total number of legal configurations placed in the reservoir at the end of the annealing step would be larger than the population size, $R$. The factor of 1.5 in the subsampling step was added as an additional safety to ensure that the reservoir is always larger than the total population. This factor acts as a tunable parameter that determines the sampling rate of the entire population. If we choose our factor so that the reservoir is nearly exactly $R$, then the weight of each member of the population is effectively measured once and there will be no benefit in comparison with standard PA. If we choose a large reservoir, many times the size of $R$, then the weight of each replica will effectively be measured many times resulting in reduced statistical errors, but at the expense of using more computer memory.
        
        \begin{figure}
            \centering
            \includegraphics[width=0.25\textwidth]{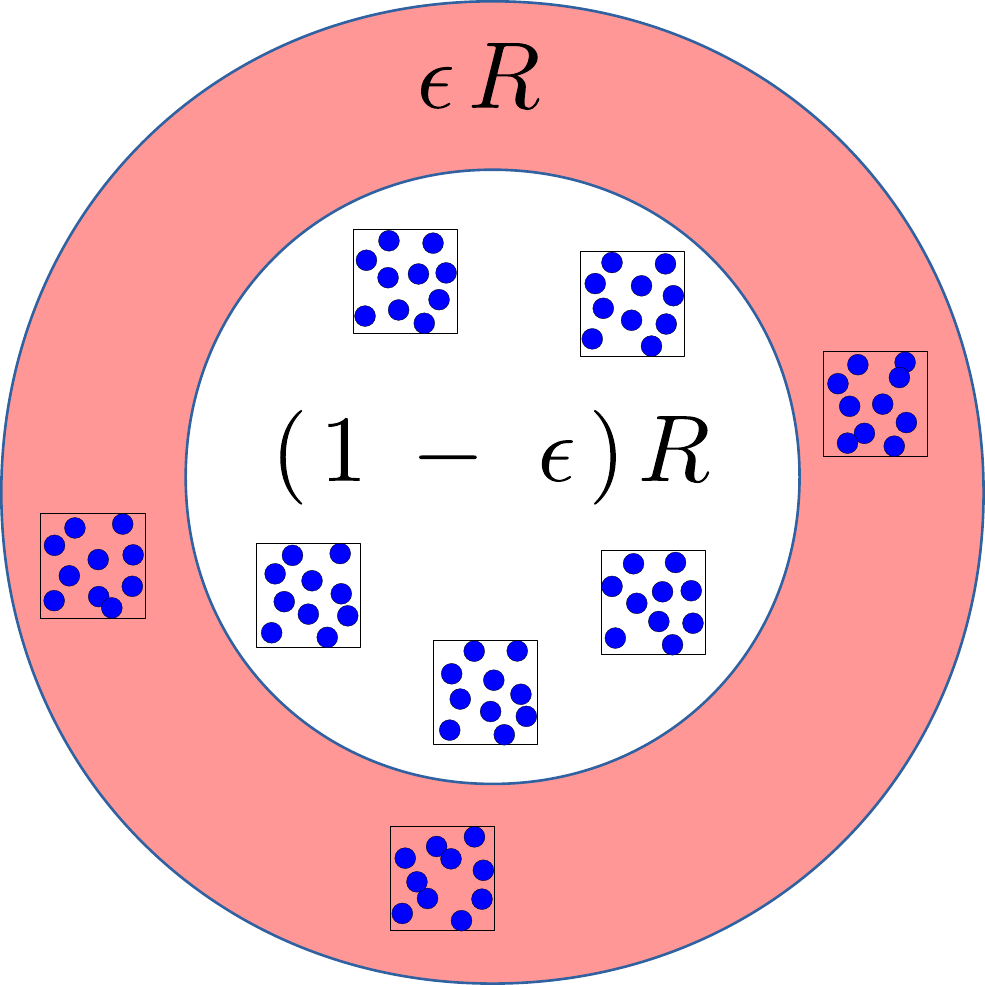}
            \caption{The outer circle corresponds to configuration space at the initial packing fraction $\phi_t$ and the inner circle corresponds to configuration space after annealing to the new packing fraction $\phi_{t+1}$. The fraction of configurations that are eliminated after annealing is an estimator for the fraction of configuration space volume that has been eliminated.}
            \label{fig:phase_space_contract}
        \end{figure}
        
        Both the hybrid and the standard versions of microcanonical PA give access to the culling fraction, $\epsilon_i$, at annealing step $i$, which is an estimator of how much configuration space volume contracts after an annealing step. In particular, the volume of configuration space decreases by a factor of $1-\epsilon_i$ each annealing step, as shown in Fig.\ \ref{fig:phase_space_contract}, and the corresponding change in entropy is given by
        \begin{align}
            \Delta S_i = \frac{1}{N}\log(1-\epsilon_i) - \log(\p_i/\p_{i-1}),
        \end{align}
        where the ratio of packing fractions corrects for the fact that we keep the system volume fixed during annealing. By summing the changes in entropy over the entire simulation, it is possible to numerically integrate the total entropy,
        \begin{align}
            \label{eq:PA_entropy}
            \Stot(\p_k) = \Stot(\p_0) + \frac{1}{N} \sum_{i=0}^{k-1}\left[ \log (1- \e_i) - \log(\p_i/\p_{i-1}) \right],
        \end{align}
        where $\Stot(\p_0)$ is the entropy at the initial packing fraction, see App.\ \ref{app:entropy_norm} for details. 
        
        Due to limitations in computational resources, it was necessary to carry out multiple independent runs of PA rather than one run with a very large population.  Results from independent simulations can be combined using weighted averaging~\cite{Wang2015,Callaham} to reduce both statistical and systematic errors and also to estimate the magnitude of these errors.  Given $M$ independent runs of PA, each with population size $R^{(m)}$, the weighted average $\overline{{\matho}}$ of an observable ${\matho}$, such as the pressure, is given by,
        \begin{equation}
        \label{eq:weight}
        \overline{{\matho}} = \frac{\sum_{m=1}^M   \tilde{\matho}^{(m)} R^{(m)} \exp[N\Stot^{(m)}] } {\sum_{m=1}^M R^{(m)}  \exp [N\Stot^{(m)} ]},
        \end{equation}
        where $\tilde{\matho}^{(m)}$ and $\Stot^{(m)}$ are the estimators of the observable and the entropy, respectively, in run $m$.  The weighted average of the entropy itself depends on a summation over the annealing schedule and is given by a different formula,
        \begin{equation}
        \label{eq:weightedS}
        \overline{\Stot} = \frac{1}{N}\log \frac{\sum_{m=1}^M R^{(m)} \exp[N \Stot^{(m)}]}{\sum_{m=1}^M R^{(m)}} .
        \end{equation}
        For fixed population size, $R^{(m)}= R$, the weighted average of an observable becomes exact in the limit of infinitely many runs, $M\rightarrow\infty$.
        
    \subsection{Event chain Monte Carlo}
        The population annealing equilibrating procedure that we use in this work is event chain Monte Carlo (ECMC), which is particularly efficient at sampling 2D and 3D hard sphere configurations~\cite{Krauth:2009, Isobe:2015, Engel:2013, Michel2014}. In the version of ECMC used here, a particle is randomly chosen and is translated in a random direction until it collides with another particle. When a collision occurs, the moving particle is stopped and the struck particle is moved in the same direction until it collides with another particle. This process is repeated until the total distance travelled by the particles, called the chain length, is equal to a predetermined length, $\ell$. When this distance is reached, the current moving particle is immediately stopped. This process is then repeated by randomly choosing a new starting particle to move in a new direction. We simulate a system with periodic boundary conditions, so it is sufficient to move particles in only the positive $x$, $y$, or $z$ directions, which violates detailed balance but preserves global balance~\cite{Krauth:2009}. 
        
        Event chain Monte Carlo provides a way of measuring the dimensionless pressure, as shown in Ref.~\cite{Michel2014}. Consider a single chain in the $x$ direction. When two particles, $j$ and $k$, collide, the distance between their centers projected in the $x$-direction is $x_k - x_j$. The ``lifted'' distance of an event chain, $x_\mathrm{final} - x_\mathrm{initial}$ is defined as
        \begin{align}
            x_\mathrm{final} - x_\mathrm{initial} = \ell + \sum_{k,j} \left(x_k - x_j \right),
        \end{align}
        where the sum takes place over all collisions in a single chain. If this process is repeated, then the dimensionless pressure is given by an average of the lifted distance over all of the chains,
        \begin{align}
            Z = \left< \frac{x_\mathrm{final} - x_\mathrm{initial}}{\ell} \right>_\mathrm{chains}.
        \end{align}
        The total entropy of the fluid is then given by the thermodynamic integral of the population-averaged dimensionless pressure,
        \begin{align}
            \label{eq:stot}
            S(\phi) = S(\phi_0) - \int_{\phi_0}^{\phi} \frac{1}{R} \sum_{r=1}^R \frac{Z_r}{\phi^\prime} d \phi^\prime,
        \end{align}
        where $Z_r$ is the dimensionless pressure of replica $r$ at packing fraction $\phi^\prime$.
        
    \subsection{Vibrational entropy}
        \label{sec:vibrational}
        There are many ways to measure configurational and vibrational entropy, see Ref.\ \cite{berthier_2019} for a review. In this section, we focus on measuring the vibrational entropies of glass states directly using two new techniques. We call the first technique the ``shell'' method, which is similar to Frenkel-Ladd thermodynamic integration~\cite{Frenkel:1984} that has been previously used to measure vibrational entropies of glassy systems~\cite{Berthier:2017}. We call the second method replica thermodynamic integration (RTI), which is a new technique that integrates the entropy of individual replicas from the fluid state into the glass state. 
        
    	\subsubsection{Constraining shell integration method}
    		\begin{figure*}[ht]
        		\includegraphics[width=0.95\textwidth]{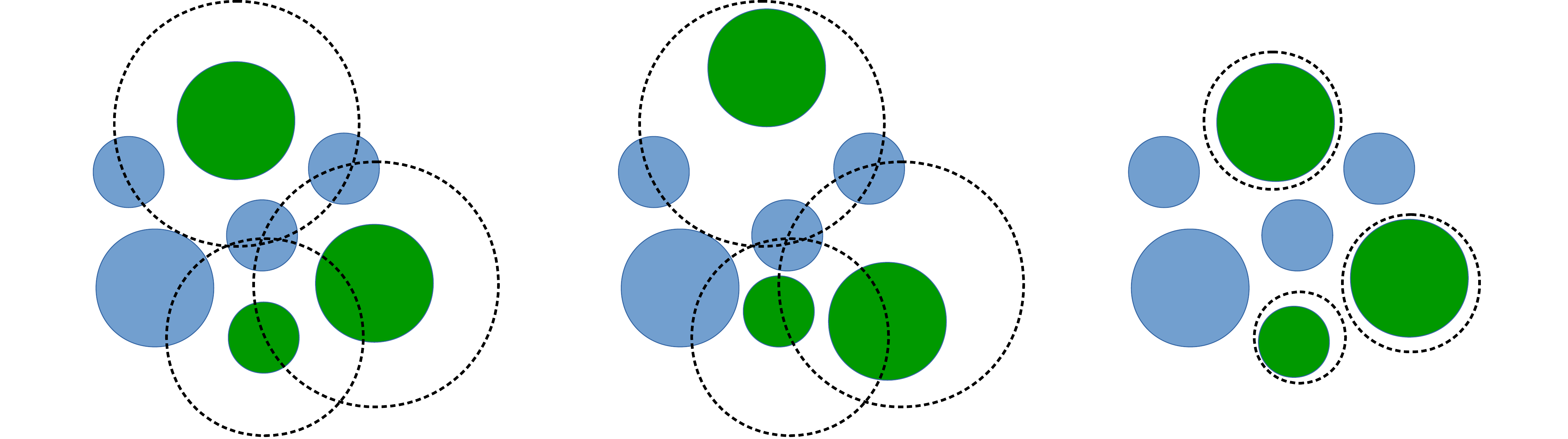}
        		\caption{Diagram representing the shell integration method. For clarity, the shells are  shown only for the three green spheres. Each sphere is contained within a hard shell that it cannot penetrate. Spheres only interact with their own shells or with other spheres that enter their shell. The shells begin very large (left) and, as the spheres are dynamically evolved the Metropolis algorithm (middle), the shells are decreased in size. Each decrement in size results in a culling in the population which corresponds to a loss of vibrational entropy. The shells are contracted and the vibrational entropy is numerically integrated via population annealing until shells no longer overlap (right). When the shells no longer overlap, the remaining entropy can be calculated analytically as an ideal gas.
                }
        		\label{fig:constrained}
            \end{figure*}
    		For the shell method, we measure the vibrational entropy of a glass state by taking the initial position of each particle as a reference. A spherical hard shell centered at each particle's reference position constrains that particle to remain within the shell. Particles are unable to penetrate their own shells, but are able to freely penetrate the shells of other particles. Initially the shells are much larger than the system size and, as the simulation progresses, they are gradually shrunk. Throughout this process, the entropy lost during each decrement in shell size is summed. Eventually all of the shells become sufficiently small that they no longer overlap with each other and only contain their own  particles. At this point, the particles can no longer interact with each other and the remaining entropy is simply that of each particle within its own shell, as seen in Fig.\ \ref{fig:constrained}. A single particle constrained within a hard shell is simply a particle in a box or, equivalently, an ideal gas.

    		The shell vibrational entropy for a given configuration of particles, $\Sshell(\vec{\pmb{x}})$, where $\vec{\pmb{x}}$ is the list of initial particle position vectors, is given by
    		\begin{align}
    		\label{eq:shell}
    		    \Sshell(\vec{\pmb{x}}) = \int_{0}^{\eta_f} \frac{d S[\pmb{r}(\eta)]}{d\eta} d\eta + K[\pmb{r}(\eta_f)],
    		\end{align}
    		where $\eta$ is a parameter that controls the shell sizes, $\eta_f$ is the parameter value where no shells overlap,  $\pmb{r}(\eta)$ is a list of the radii of all of the shells, and $K[\pmb{r}(\eta_f)]$ is the sum of the ideal gas entropies of each particle in its shell. During the integration, all shells shrink at the same rate. 
    		When a shell no longer overlaps with other shells, then its integration stops, its shell radius no longer shrinks, and the sphere contained within the shell no longer contributes to the numerical integral. The remaining vibrational entropy for the sphere/shell pair is calculated analytically and contributes to the constant $K$. The integration continues with the remaining shells that have overlaps. When no more shells overlap, the integration is complete and 
    		\begin{align}
    		    \label{eq:s_id}
    		    K[\pmb{r}(\eta_f)] = \frac{1}{N} \sum_{i} \log\left[ \frac{4\pi}{3}(r_\mathrm{shell}^i - r_\mathrm{sphere}^i)^3 \right] + \frac{3}{2},
    		\end{align}
    		where $i$ enumerates the particles and $r^i_{\mathrm{shell}}$ is the final radius of the $i^{\rm th}$ shell. This formula is straightforward to understand, $\frac{4\pi}{3}(r_\mathrm{shell}^i - r_\mathrm{sphere}^i)^3$ is simply the free volume of each shell and $3/2$ is the kinetic contribution to the entropy of the ideal gas. In general, an additional factor of $-3\log(\lambda_\mathrm{th})$ is present, where $\lambda_\mathrm{th}$ is the thermal deBroglie wavelength, but, as before, we set $\lambda_\mathrm{th}=1$.
    		
    		This method of calculating $\Svib$ constrains the system so that particles are unable to switch places and, therefore, are distinguishable. This approximation becomes more accurate as structural relaxation times become very large deep within the glassy regime. This method may also be used to measure the total entropy in the fluid regime, however, since particles are no longer localized in cages, an additional term of $1-\log(N/2)$ must be included in order to account for indistinguishability of the particles. 
    		
    		The numerical integration of $\Sshell(\phi)$ is performed using population annealing as follows:
    		\begin{enumerate}
    		    \item Choose a random glass sample from an equilibrium distribution at fixed $\phi$
    		    \item Make $\Rs$ identical copies of the glass sample
    		    \item Initialize shells so they are infinitely large
    		    \item Evolve population using MD or MC
    		    \item Anneal in shell size and integrate entropy.
    		\end{enumerate}
            The value of $\Sshell$ will depend on the number of sweeps per annealing step (see Fig.\ \ref{fig:log_entropy}) leading to some ambiguity in the definition of $\Svib$ and $\Sc$. 
            The details of the parameters used for the $\Sshell$ integration will be described in detail in Sec.\ \ref{sec:parameters}.
    		
    	\subsubsection{Replica thermodynamic integration method}
    	    The second method of measuring the vibrational entropy is called replica thermodynamic integration (RTI). 
    	    In abstract terms, the RTI method can be described by recasting Eq.\ \ref{eq:svib} into a thermodynamic integral,
        	\begin{align}
        		\Svib(\phi) = S(\phi_0) - \sum_{\nu\in c(\phi)} w_\nu(\phi) \int_{\phi_0}^{\phi} \frac{Z_\nu(\phi^\prime)}{\phi^\prime}\, d\phi^\prime + C.
        	\end{align}
        	where $Z_\nu(\phi^\prime)$ is an equilibrium trajectory of pressures of increasing density that ends in glass state $\nu$ at packing fraction $\phi$, $w_\nu(\phi)$ the weight of $\nu$ at packing fraction $\phi$, and $C$ is a normalizing constant. This expression is different from the thermodynamic integral for the total entropy which is given by
            \begin{align}
        		S(\phi) = S(\phi_0) - \int_{\phi_0}^{\phi} \sum_{\nu\in c(\phi^\prime)} w_\nu(\phi^\prime) \frac{Z_\nu(\phi^\prime)}{\phi^\prime}\, d\phi^\prime.
        	\end{align}
        	In the fluid phase, for $\phi < \phid$, there is only one thermodynamic state and the two expressions are identical, however, in the glassy phase, the two become distinct due to the formation of ergodically separate glass states.
    	    
    	    The RTI method can be realized in a PA simulation by integrating the entropy along the history of replica $r$ at packing fraction $\p$, denoted as $S_r(\p)$. Using the ECMC dimensionless pressure along the history,
             \begin{align}
                S_r(\phi) = S(\phi_0) - \int_{\phi_0}^{\phi} \frac{Z_{\kappa(\phi^\prime|r,\phi)}}{\phi^\prime} d \phi^\prime ,
            \end{align}
            where $\kappa(\phi^\prime|r,\phi)$ is the replica index of the ancestor at packing fraction $\phi^\prime$ of replica $r$ at  packing fraction $\phi$ and $S(\phi_0)$ is the initial entropy at the low packing fraction $\phi_0$.  In PA, the population is an equilibrium sample of the glass states that automatically takes into account weights during resampling. 
            This means that  we can simply average  $S_r$ over the population to get the average glass entropy $\Srti(\p)$, albeit with an incorrect constant of integration set in the low density limit,  
            \begin{align}
                \Srti(\p) = \frac{1}{R}\sum_{r=1}^R S_r(\phi).
            \end{align}
            In the fluid regime, where replicas are individually in equilibrium, the two quantities are equal. In the glassy regime, $\Srti>\Stot$ (see Fig.\ \ref{fig:Sc_unnorm}) because $\Srti$ only averages over the surviving, high-entropy replicas whereas $\Stot$ averages over all replicas.
            
            To identify the  correct constant of integration requires us to use the shell method or an equivalent approach.  We define $\Srti^{(\p')}(\p)$ with an additive constant so that it is equal to $\Sshell(\p')$ at packing fraction $\p'$,  
            \begin{align}
            \label{eq:srtiprime}
                \Srti^{(\p')}(\p) =  \Srti(\p) + \left[\Sshell(\p')-\Srti(\p')\right].
            \end{align}
    	   $\Srti^{(\p')}(\p)$ provides quasi-continuous estimate of $\Svib(\p)$ if the constant of integration is set at an appropriate point $\p'>\phid$ where $\Sshell$ can be reliably measured. It is worth mentioning that $\Srti$ measures entropy in a way where particle swaps are allowed and therefore naturally includes mixing entropy.
    
    \subsection{Equilibration} \label{sec:micro_PA_error}
        The equilibration of our simulations can be estimated in several ways. The first and perhaps most obvious is to use one of the intrinsic equilibration metric associated with population annealing~\cite{Wang2015}.  A second method, discussed at the end of this section, is specific to glasses and depends on the configurational entropy.
        
        It can be shown~\cite{Wang2015, Callaham} that systematic errors in PA scale as $1/R$ and that a prefactor of this scaling is an ``equilibration population size", $\rhof$,
        \begin{align}
            \rhof = \lim_{R \rightarrow \infty} R\, \var(N\Stot),
        \end{align}
        where $R$ is the population size of the simulation and $\Stot$ is the total entropy estimator, and the variance is measured with respect to independent runs of PA.  In practice, $\rhof$ must be estimated from multiple runs with finite but sufficiently large population size. The intuition behind relating the variance of $\Stot$ to systematic errors is that when the entropy estimator has large fluctuations then independent simulations sample distinct regions of configuration space and produce different results.  Furthermore, these results are, on average, biased since the correct value of an observable would be obtained from an entropically weighted average over a very large number of runs. 
        
        Although results from several independent simulations can be combined using weighted averaging, $\rhof$ is not a suitable measure of equilibration of the weighted average of many simulations. The extension to multiple runs was introduced in Ref.~\cite{Callaham}, using the quantity $\rhofs$, 
        \begin{align} 
            \label{eq:rhofs}
            \rhofs = \lim_{M\rightarrow \infty}  R_\mathrm{tot}\, \var(N\overline{\Stot}),
        \end{align}
        where $M$ is the number of simulations, $\overline{\Stot}$ is defined in Eq.\ \ref{eq:weightedS} and
        \begin{align}
            R_\mathrm{tot}=\sum_{m=1}^M R^{(m)},
        \end{align}
        is the total population size of all the simulations, with $R^{(m)}$ the population size of the $m^{\rm th}$ simulation. 
        It can be shown that $\rhofs /R_\mathrm{tot}$ is proportional to the systematic errors in observables obtained from the weighted average of multiple runs, all carried out with the same annealing schedule but with, perhaps, different population sizes. 
        Although $\rhofs$ was originally calculated using bootstrap resampling~\cite{Callaham}, we found that it is preferable to estimate $\var(\overline{\Stot})$ using a weighted variance of the entropies of the runs,
        \begin{align}
            \var(N\overline{\Stot}) \approx \frac{N^2 \sum_{m=1}^M R^{(m)} e^{N\Stot^{(m)}} \left(\Stot^{(m)} - \overline{\Stot}\right)^2}{\sum_{m=1}^M R^{(m)} e^{N\Stot^{(m)}}}.
        \end{align}
        This method gives nearly identical results to the bootstraps method used in Ref.~\cite{Callaham}, but is easier to calculate.
        
        We can also use the configurational entropy to estimate whether the population size of the simulations is sufficient. The change in configurational entropy determines how the number of glass states decreases as the density is increased and can be used to estimate the rate of die-off of independent population members in PA. If the glass transition is suitably sharp, then it is safe to assume that the population is in equilibrium before the dynamic transition and therefore, at $\phid$, the number of independent glass states is approximately equal to the population size, $R$. Since the transition is relatively sharp, it is also safe to assume that no new glass states will be discovered after the dynamic transition and that population members remain stuck in the same glass state as their ancestor at $\phid$. With this in mind, the number of independent glass states in our population for $\phi > \phid$, $N_\mathrm{g}$, is
        \begin{align}
            N_\mathrm{g}(\phi) = R e^{N \left[\Sc(\phi) - \Sc(\phid) \right]}.
        \end{align}
        This means that in order to have multiple independent glass states at packing fraction $\phi$, the population size must satisfy
        \begin{align}
            \label{eq:exp_delta_s}
            R \gg e^{-N \Delta \Sc},
        \end{align}
        where $\Delta \Sc = \Sc(\phi) - \Sc(\phid)$. The implication of this is that PA can only go a short way beyond the dynamic transition specific to the MCMC used to equilibrate the population. 

    \subsection{Simulation details} \label{sec:parameters}
    	\begin{table}
            \centering
    		\setlength{\tabcolsep}{8pt}
    		\begin{tabular}{| l | l | l | l | l | l |}
    		    \hline
    			$N$ & $R$ & Sweeps & $M$ & $\epsilon$ & $\p_f$\\
                \hline
                \multirow{2}{*}{30} & $ 10^6$ & $2.3 \times 10^5$ & 60 & \multirow{2}{*}{0.074} & 0.63 \\
                & $10^5$  & $2.2 \times 10^6$ & 10 & & 0.625\\
                \hline
                \multirow{2}{*}{60} & $3 \times 10^6$ & \multirow{2}{*}{$1.4\times 10^{5}$} & 8 & \multirow{2}{*}{0.15} & \multirow{2}{*}{0.625} \\
                & $5 \times 10^6$  &  & 14 & &\\
                \hline
                100 & $10^6$ & $1.0\times 10^{5}$ & 10 & 0.12 & 0.625 \\
    			\hline  
    		\end{tabular}
    		\caption{Parameters for equilibrium population annealing runs: $N$ is the number of particles, $R$ is the population size of each run, Sweeps is the total number of ECMC sweeps per replica per run, $M$ is the number of independent simulations, $\epsilon$ is the average culling fraction, and $\p_f$ is the highest packing fraction in the runs. For $N=60$, there were two population sizes, $3\times 10^6$ and $5\times10^6$, whose simulations were combined using weighted averaging. For $N=30$, the two sets of simulations were analyzed separately.
    		}
    		\label{table:equilibrium}
    	\end{table}
    	We ran two different sets of large-scale simulations. The main set of simulations produced ensembles of glass states and measured  values of $\Stot$, $\Srti$, and $Z$ for a quasicontinuous set of packing fractions up to packing fraction $\p_f$ in the glassy regime. The second set of simulations measured $\Sshell$ for many glass samples in order to obtain an equilibrium measure of the vibrational entropy at different packing fractions and to normalize $\Srti$.
    	
         The parameters used in the first simulation are shown in Table\ \ref{table:equilibrium}. The main simulations began at $\phi_0=0.3$, where it is still possible to efficiently sample configurations by randomly placing spheres in a box using a rejection method. The entropy is normalized relative to the ideal gas at $\phi=0$ with $T=1$ and $\lambda_\mathrm{th}=1$, see App.\ \ref{app:entropy_norm}. To get equilibrium samples at higher densities, we ran population annealing Monte Carlo (PA) with Event Chain Monte Carlo (ECMC) as the equilibrating dynamics. Our ECMC simuluations had chain length equal to a fixed fraction, 0.618, of the box size.  This choice of chain length is significantly longer than that of Ref.\ \cite{Callaham}. In retrospect, the choice in Ref.\ \cite{Callaham} would have been preferable. 
        For most of the simulations, the chain schedule is given by,
    	\begin{align}
    	    \label{eq:schedule}
    	    \text{\# event chains} = 
            \begin{cases}
                20 &\phi \leq 0.54 \\
                60 &0.54 < \phi \leq 0.59 \\
                10 & \phi >0.59. \\
            \end{cases}
    	\end{align}
    	The idea is to do enough chains to equilibrate in the fluid regime and partially equilibrate near the dynamic transition. At $\phi>0.59$, when the simulation is beyond the dynamic transition, we no longer depend on using ECMC to equilibrate the entire population. Instead, we only perform a few ECMC sweeps in an attempt to equilibrate each population member locally within its own glass state. For $N=100$, the number of event chains at each annealing step was halved, but the number of annealing steps was doubled so that the number of chains per change in packing fraction was kept constant. This was done so that the $N=100$ culling fraction remained moderate. We also performed 10 simulations with $N=30$, $R=10^5$ and $\times 10$ the number of chains as in Eq.\ \ref{eq:schedule}, see Table\ \ref{table:equilibrium} for details.
    	
        The parameters used in the shell vibrational entropy simulations are presented in Table\ \ref{table:vibrational}. We are able to measure the vibrational entropy of individual glass states with relatively small population sizes, $\Rs$, because the motion in configuration space is confined within a single glass state. 
        Larger population sizes were tried as well and were found to produce numerically identical results which are not presented here. In order to equilibrate after each annealing step, we use the Metropolis algorithm with a fixed step size proposal chosen at each annealing step such that the acceptance ratio is always between 40\% and 45\%. The constraining shells are chosen such that their initial size is larger than the box containing the particles and are shrunk so that 10\% of the population is culled at each annealing step. 
        Error bars for the shell method are obtained by bootstrapping over all of the shell integrated vibrational entropies.
    
    	\begin{table}
    	    \centering
    		\setlength{\tabcolsep}{8pt}
    		\begin{tabular}{| l | l | l | l | l |}
    		    \hline
    			$N$ & $\Rs$ & Steps & Sweeps/step & Samples \\
    			\hline
                30 & $10^4$ & 3018 & 1600 & 300 \\
                \hline
                60 & $3\times10^4$ & 6056 & 800 & 110 \\
    			\hline  
    		\end{tabular}
    		\caption{Simulation parameters for vibrational entropy measurements using the shell method. The number of steps was different for $N=30$ and $N=60$ in order to keep the rate of shell contraction constant.
    		}
    		\label{table:vibrational}
    	\end{table}

\section{Results} \label{sec:results}
    In this section we present results from a large-scale computational study of the binary hard sphere system using population annealing Monte Carlo to sample equilibrium glass states at high density. We begin by presenting equilibrium measurements of the dimensionless pressure and the entropy as a function of packing fraction in Sec.\ \ref{sec:results_Z_S}. Following this, we present our measurements of the configurational entropy and a detailed comparison between the two methods used to estimate the vibrational entropy in Sec.\ \ref{sec:results_Sc_Svib}. We then compare the locations of the estimated jamming density, $\phij$, and Kauzmann transition location, $\phik$, in Sec.\ \ref{sec:results_transitions}. Finally, we present several metrics to assess the equilibration of the simulations in Sec.\ \ref{sec:results_equilib}.

	\subsection{Pressure and total entropy}
	    \label{sec:results_Z_S}
	    
		\begin{figure}
            \centering
    		\includegraphics[width=0.45\textwidth]{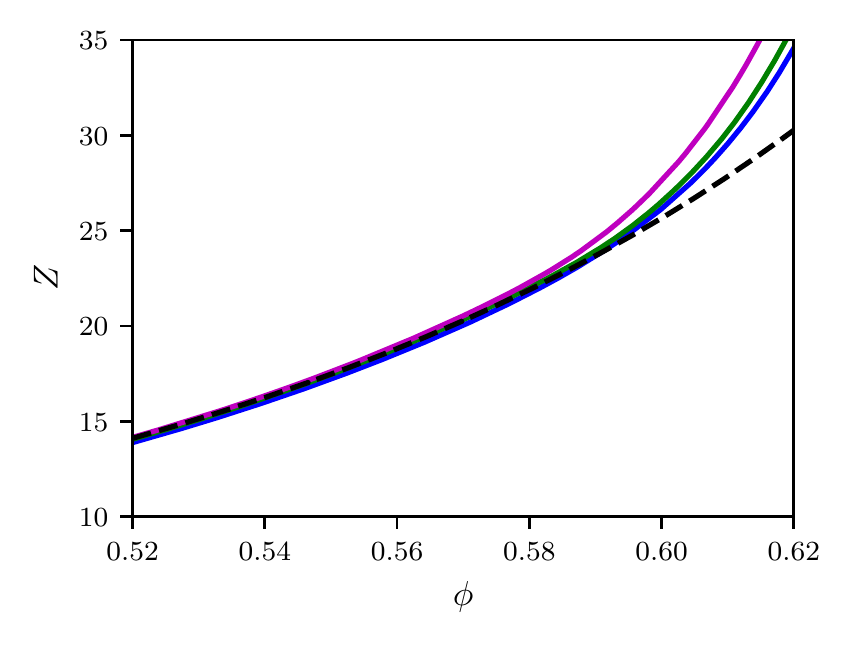}
			\caption{The dimensionless pressure, $Z$, as a function of packing fraction, $\phi$ for system sizes 30 (blue, bottom), 60 (green, middle), and 100 (purple, top). The dashed line is the phenomenological BMCSL equation of state (Eq.\ \eqref{eq:BMCSL}).} 
			\label{fig:EOS}
		\end{figure}
		
    	The dimensionless pressure, $Z$, is shown in Fig.\ \ref{fig:EOS}, where the solid lines correspond to simulations and the dashed black line corresponds to the BMCSL equation of state. Our simulation results deviate from the phenomenological equation of state at high packing fractions after the dynamic glass transition. 
        Given that our simulations for sizes $N=30$ and $N=60$ are believed to be in statistical equilibrium for values of $\phi$ past $\phid$ and that the BMCSL equation of state is not correct for high density, as is clear by its divergence at $\phi=1$, the deviation is at least partially due to the proximity to the true divergence of the pressure at random close packing, see Sec.\ \ref{sec:results_transitions}.
    	
		\begin{table}
            \centering
			\begin{tabular}{| c | l | l | l | l | l | l| }
			    \cline{2-7}
			    \multicolumn{1}{c|}{} & \multicolumn{3}{c |}{$Z$} & \multicolumn{3}{c |}{$S$} \\
			    \hline
				$\phi$ & {\footnotesize this work} &  {\footnotesize Ref.~\cite{Callaham}} &  {\footnotesize BMCSL}  & {\footnotesize this work} & {\footnotesize Ref.~\cite{Callaham}} & {\footnotesize BMCSL} \\
				\hline
				0.58 & 22.01(1) & 22.04(3) & 21.90 & -1.230 & -1.210 & -1.240\\
				\hline
				0.59 & 23.90(6) & 23.99(14) & 23.69 & -1.621 & -1.603 & -1.629 \\
				\hline
				0.60 & 26.4(1) & 26.5 & 25.66 & -2.044 & -2.026 & -2.043 \\
				\hline  
			\end{tabular}
			\caption{
			The dimensionless pressure and total entropy for $N=60$ at several values of $\phi$ from this work compared to values obtained from Ref.~\cite{Callaham} and to the phenomenological BMCSL equation of state (Eq.\ \ref{eq:BMCSL}). Reference\ \cite{Callaham} entropies were modified so as to be consistent with the normalization used in the present work and to correct for the log volume term in Eq.\ \ref{eq:PA_entropy}, missing from that reference.}
			\label{table:results}
		\end{table}

        In addition to the pressure, we have also measured the total entropy per particle, $\Stot$, using ECMC, Eq.\ \ref{eq:stot}. For $N=60$, these results can be seen in comparison to measurements from previous works in Table.\ \ref{table:results}.
        
    \subsection{Vibrational and configurational entropies}
        \label{sec:results_Sc_Svib}
        \begin{figure}
            \centering
    		\includegraphics[width=0.45\textwidth]{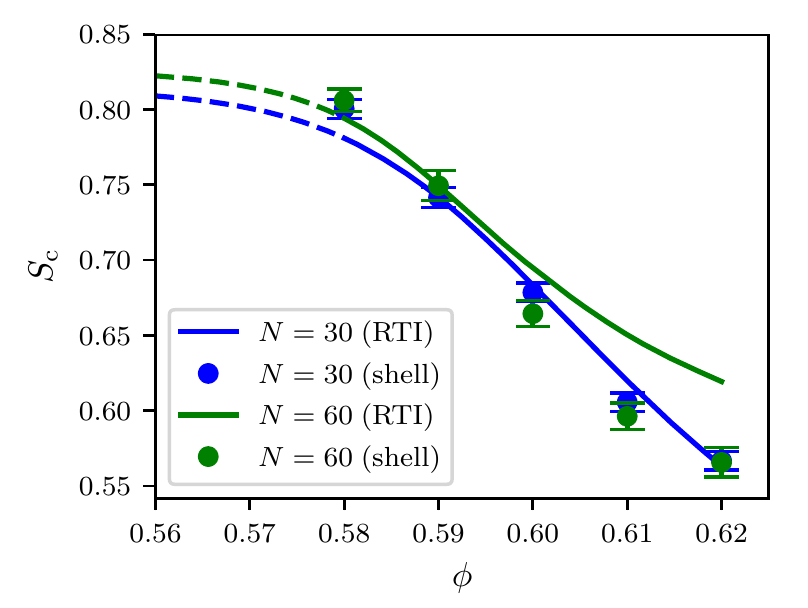}
    		\caption{The configurational entropy measured via the replica thermodynamic integration method (lines) and the shell integration method (dots) for $N=30$ (blue, lower) and $N=60$ (green, upper lines). The dashed line corresponds to values of $\phi$ that are below the dynamic transition where the system behaves as a fluid. The RTI method produces a value of $\Sc$ at every annealing step, while the shell method was performed at $\phi= 0.58$, 0.59, 0.60, 0.61, and 0.62.
    		} 
    		\label{fig:Sc}
    	\end{figure}
        We measured the configurational entropy using two different methods to obtain the vibrational entropy: replica thermodynamic integration (RTI) and shell integration. The results for these two methods are shown in Fig.\ \ref{fig:Sc}, where the continuous curves correspond to the RTI method and the points correspond to the shell method. The constant of integration for RTI is obtained from the shell method at $\p=0.59$, that is, we use $\Srti^{(0.59)}$.

        For the $N=30$ system, the two methods are consistent with each other for the entire range of the simulation while for $N=60$, the two methods agree for $\phi \lesssim 0.595$. The difference between the two methods for $N=60$ and $\phi > 0.595$ may be due to an inadequate number of sweeps for the RTI method to estimate $\Svib$.
        
	    \begin{figure}
            \centering
    		\includegraphics[width=0.45\textwidth]{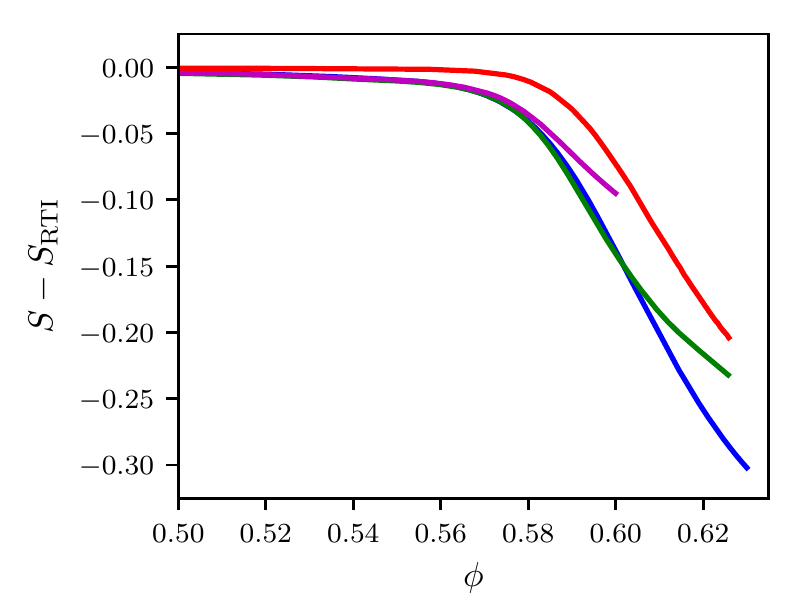}
    		\caption{$\Stot-\Srti$
    		as a function of $\p$ for $N=30$ (blue, bottom), $N=60$ (green, second from bottom), and $N=100$ (purple, second from top) with the simulation parameters of Table \ref{table:equilibrium}. The red curve (top) corresponds to an $N=30$ simulation with ten times the number of Monte Carlo sweeps and exhibits a sharper and slightly later dynamic transition than the other simulations.
    	    }
    		\label{fig:Sc_unnorm}
    	\end{figure}
    	
        Although equilibrium quantities such as $Z$ and $S$ are largely independent of the annealing schedule provided sufficiently large population sizes are used, the values of $\Srti$ and $\phid$, defined roughly as the location of the shoulder of the  $\Stot-\Srti$ curve, depend on the annealing schedule. This can be seen explicitly in Fig.\ \ref{fig:Sc_unnorm}, where the blue, green, and purple curves correspond to $\Stot-\Srti$ measurements using the parameters of Table \ref{table:equilibrium} for simulations of $N=30$, 60, and 100 particles, respectively. The blue, green and purple curves come from simulations with the same number of ECMC chains per unit change in packing fraction. The outlying red curve uppermost in the plot corresponds to an $N=30$ simulation with $\times 10$ the number of ECMC chains as that of the standard $N=30$ simulation (see Eq.\ \ref{eq:schedule}).  Here $\phid$ is significantly increased due to the increased ECMC equilibration. For $\p>\phid$ the curve is nearly parallel to the other curves and, after setting the constant of integration using the shell method, the configurational entropy obtained from this high sweep number simulation is nearly the same as obtained in the main, lower sweep number simulations.
        
        Finally, Fig.\ \ref{fig:Sc_unnorm} illustrates an important feature of population annealing.  For $\p<\phid$, we see that $\Stot-\Srti$ is nearly zero.  This is because each replica is equilibrated by ECMC and has the same properties as the ensemble of replicas. On the other hand, for $\p>\phid$ ergodicity is broken and each replica is confined to a single glass state so that the ensemble average no longer equals an average obtained from a single replica using ECMC.  Nonetheless, the resampling step in PA ensures that  ensemble averages represent equilibrium properties well beyond $\phid$, as discussed in more detail in Sec.\ \ref{sec:results_equilib}.   

    	\begin{figure}
            \centering
    		\includegraphics[width=0.45\textwidth]{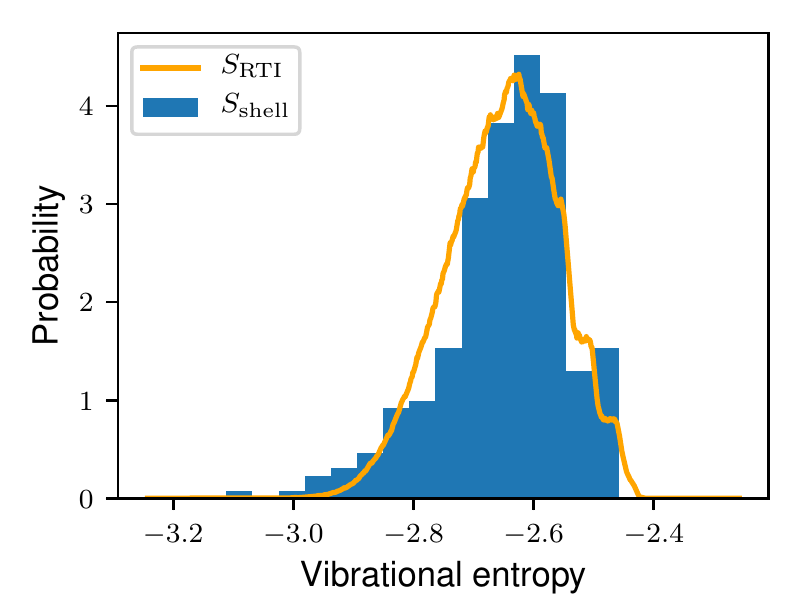}
    		\caption{Probability density function (PDF) for  $\Svib$ calculated with the shell method (histogram) and RTI method (curve) for $N=30$, $\phi=0.60$.
    	    }
    		\label{fig:svib_hist}
    	\end{figure}
    	
		\begin{table}
		    \centering
			\setlength{\tabcolsep}{8pt}
			\begin{tabular}{| l | l | l | l | l | }
			    \hline
				$\Svib$ & $\mu$ & $\sigma$ & skewness & kurtosis \\
				\hline			
				shell & -2.65 & 0.11 & -0.90 & 1.29 \\
				\hline
				RTI & -2.65 & 0.09 & -0.25 & 2.76 \\
				\hline  
			\end{tabular}
			\caption{Statistics of the probability distribution functions of the vibrational entropy measured by the shell and RTI methods for $N=30$, $\phi=0.60$.
			}
			\label{table:distributions}
		\end{table}
    	Figure\ \ref{fig:Sc} shows that values of $\Svib$ obtained from RTI and the shell method give consistent results when averaged over many glass states. 
    	However, looking at the distribution of values from the two methods we can see some interesting and, as yet, not fully explained differences.
    	Figure \ref{fig:svib_hist} shows the measured probability distribution functions of $\Sshell$ and $\Srti$ for $N=30$ at $\phi=0.60$. The histogram corresponds to $\Sshell$ values from 300 samples chosen randomly from the PA simulations and the  curve corresponds to $\Srti$ values from $6\times10^7$ samples. The data was normalized so that the averages of $\Srti$ and $\Sshell$ are equal at $\phi=0.59$, which is the same as the normalization used for $\Sc$. As seen in the plot and Table\ \ref{table:distributions}, both distributions exhibit the same general features, but there is a significant discrepancy between the two in the low entropy tail where $\Sshell$ has significantly more weight than $\Srti$. 
        Another interesting characteristic of the distributions is the presence of flat steps in the RTI distribution and a second maximum in the shell histogram at high entropies. 
    	
    	\begin{figure}
            \centering
    		\includegraphics[width=0.45\textwidth]{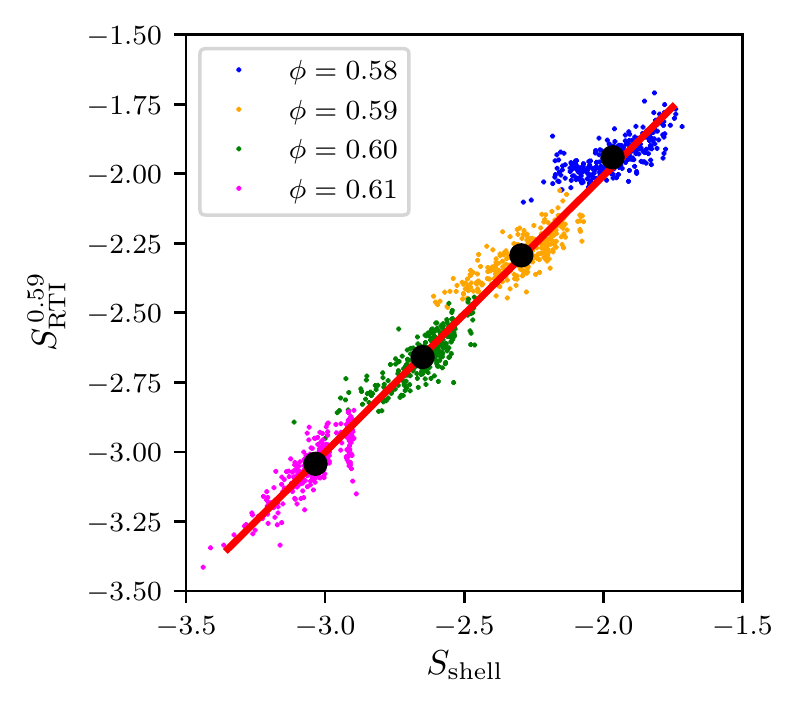}
    		\caption{Scatter plot of $\Srti^{(0.59)}$ versus $\Sshell$ for 300 $N=30$ glass samples at $\phi= 0.58$, 0.59, 0.60, and 0.61 (from right to left). The line of best fit for $\phi= 0.59$, 0.60, and 0.61 has a slope of 0.99 and intercept -0.02, showing excellent correspondence. The vertical sets of high-entropy values that can be seen at $\phi=0.59$ through $0.61$ correspond to the high-entropy plateaus in the vibrational entropy histograms.
    	    }
    		\label{fig:shell_line}
    	\end{figure}
    	
        A different perspective on the relationship between $\Sshell$ and $\Srti$ is seen in their joint probability distribution, shown in Fig.\ \ref{fig:shell_line}.  The  different sets of colored points correspond to glass states at different packing fractions, the large black points correspond to the average entropies at each packing fraction, and the red line is a line of best fit for the $\phi=0.59$, 0.60, and 0.61 data. At high packing fractions, $\Sshell$ and $\Srti$ are strongly correlated and the resulting best fit line has slope 0.99 and intercept -0.02. At these densities, the joint distributions are approximately symmetric about the linear fit, but there are two distinct features of the joint distribution that require explanation. The first is that there are clusters of points at high entropy that have a  narrow distribution of $\Sshell$ values and a wide distribution of $\Srti$ values.  These clusters are most pronounced for high packing fraction. The second feature is that the low entropy tail of the distribution is skewed toward higher values of $\Srti$.  This feature is most pronounced for low packing fraction. The high entropy clusters  correspond to the rightmost plateau in the histogram shown in Fig.\ \ref{fig:svib_hist}. These high entropy clusters require more investigation.

        The points in the low entropy skewed part of the joint distribution that are apparent for lower packing fractions in Fig.\ \ref{fig:shell_line} may represent configurations of particles that are at least partially fluid-like having diffusive rather than caged particle motions.  While the RTI method should capture the full entropy of these configurations, the limited number of Metropolis sweeps used in the shell method may fail to fully explore the configuration space of these fluid-like configurations.  On the other hand, for high packing fraction, the shell method finds high entropy glass states for which the RTI method finds substantially lower entropies.  This may be due to the small number of ECMC chains per annealing step used in the RTI method.
        
        One possible issue with $\Sshell$ is that it is numerically poorly behaved at high density. When initializing the shells for a glass state, there is a possibility that two spheres will be nearly touching. The resulting shells will be only slightly larger than the sphere sizes and, accordingly, the ideal gas entropies associated with those shells will be large due to the logarithm of the volume present in Eq.\ \ref{eq:s_id}. The numerical integration of $\Sshell$ will compensate for the logarithmic term by having many more annealing steps, however, this amounts to subtracting one large number from another and is, therefore, error-prone. This numerical instability is inherent to Frenkel-Ladd techniques and becomes more problematic as the density approaches jamming.
        
        \begin{figure}
            \centering
    		\includegraphics[width=0.5\textwidth]{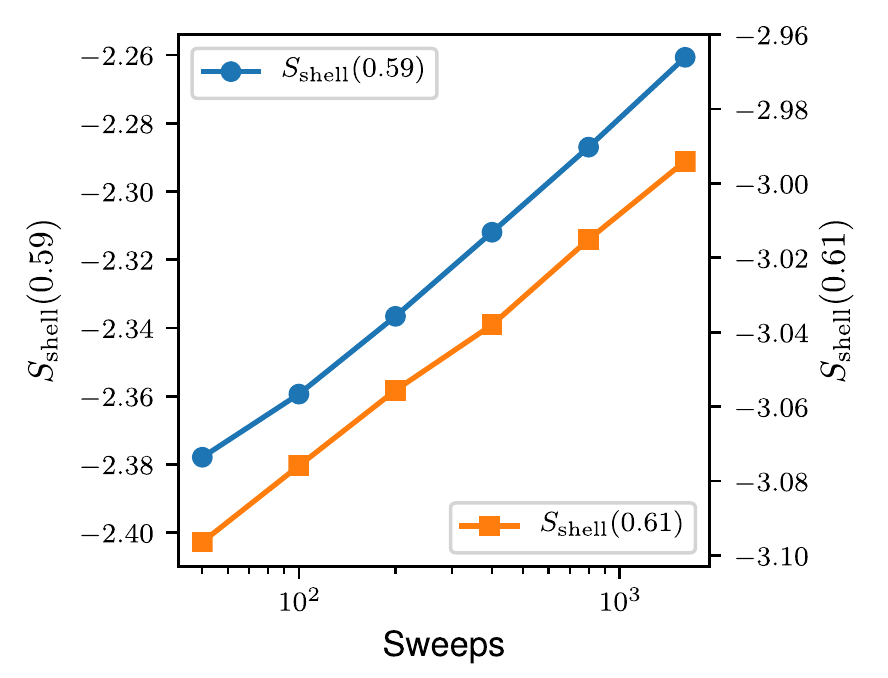}
    		\caption{Shell vibrational entropy, $\Sshell(\phi)$, as function of sweeps per annealing step at packing fractions 0.59 (circles, left axis) and 0.61 (squares, right axis) for $N=30$. Both curves show logarithmic growth with the number of sweeps.
    		}
    		\label{fig:log_entropy}
    	\end{figure}	
    	
        Vibrational entropy, and thus also configurational entropy, is not uniquely defined because glass states are metastable and vibrational entropy slowly increases as the time allowed to explore configuration space increases. Figure\ \ref{fig:log_entropy} is a plot of the shell vibrational entropy as a function of sweeps for glass samples at $\phi=0.59$ and $\phi=0.60$, where each data point corresponds to an average over twenty different samples. The plot exhibits a linear-log behavior, which is consistent with the known logarithmic relaxation dynamics of configurational glasses. 

    \subsection{Transition locations}
        \label{sec:results_transitions}
        \begin{figure*}
            \centering
            \includegraphics[width=0.45\textwidth]{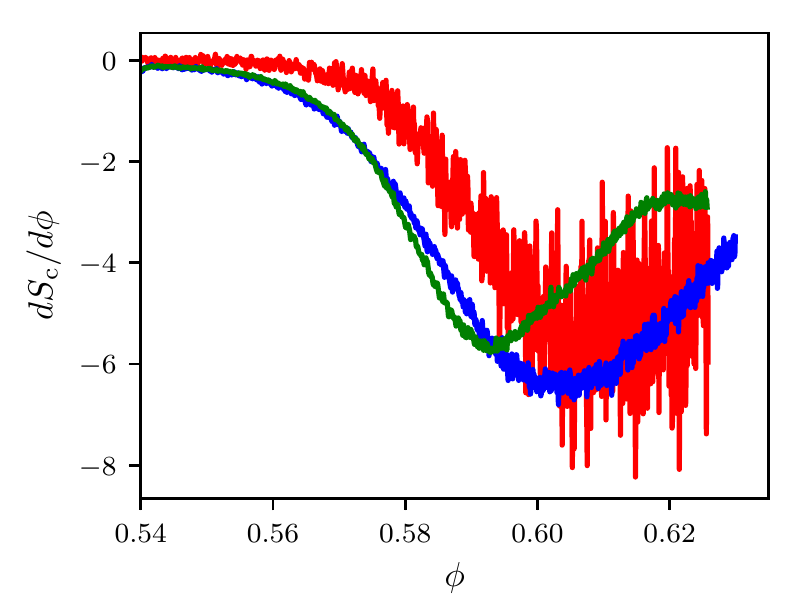}
            \includegraphics[width=0.45\textwidth]{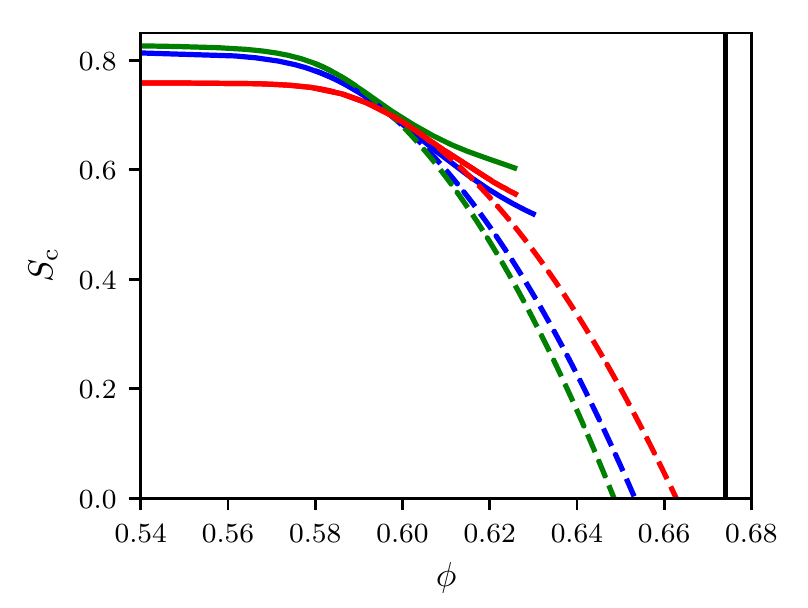}
            \caption{(left panel) The derivative of the configurational entropy with respect to packing fraction for $N=30$,  (simulations with $2\times 10^5$ sweeps, blue; simulations with $2\times 10^6$ sweeps, red)  and $N=60$ (green). We use the $d\Sc/d\phi$ plot to determine the fitting range for our data. (right panel) Solid lines are $S_c$ curves from the RTI data and dashed lines are best fit quadratic extrapolation of these curves.  For $N=30$ with $2\times10^5$ and $2\times 10^6$ sweeps, the fit is over the range $\p=0.58$ to $0.595$ and $\p=0.585$ to $0.595$, respectively, and for $N=60$ the fit is over the range $\p=0.58$ to $0.59$.  The vertical solid line is the  jamming point as determined by the free volume fit.}
            \label{fig:Sc_extrapolation}
        \end{figure*}
        
        The jamming density, $\phij$, can be estimated by making a free volume fit~\cite{salsburg:1962} to the dimensionless pressure at densities beyond equilibrium,
    	\begin{align}
    	    Z = \frac{d'\phij}{\phij-\phi}.
    	\end{align}
    	We make this fit for $\phi > 0.61$ and obtain values of $d'$ equal to 2.83 and 2.85 and $\phij$ equal to 0.676 and 0.673 for $N=30$ and $N=60$, respectively. We also performed this for $N=30$ with ten times the number of sweeps and obtained an estimated $\phij$ of 0.677 with $d'=2.91$. 
    	Because our simulations have fallen out of equilibrium over most of the range of the fits, the measured dimensionless pressures are higher than their equilibrium values and the estimated values of $\phij$ act as lower bounds to the random close packed density,  $\prcp$. Our measured values of $\phij = 0.673$ and $0.676$ are slightly larger than those found in Ref.\ \cite{OdBe11} and within error bars of those found in Ref.\ \cite{Callaham}. 
    	
    	We can estimate the location of the Kauzmann transition by extrapolating $\Sc\rightarrow0$ with our curves obtained with the RTI method. As shown in the left panel of  Fig.\ \ref{fig:Sc_extrapolation}, the $d\Sc/d\phi$ curves are linear for $\phi > \phid$ as long as the simulation remains in equilibrium. We make a quadratic fit to $\Sc$ for $N=30$ with $0.58 < \phi < 0.595$ and $N=60$ with $0.58 < \phi < 0.59$ and obtain $\phik$ estimates of 0.653 and 0.649, respectively. We also performed this for $N=30$ with ten times the number of event chains over the range of $0.585 < \phi < 0.595$ and obtained an estimated $\phik$ of 0.663.  These fits are shown as dotted lines in the right panel of Fig.\ \ref{fig:Sc_extrapolation}. The fit ranges were chosen to be after the dynamic glass transition but within the equilibrated regime. The resulting extrapolations are generally consistent and support the inequality $\phik < \phij \leq \prcp$. Note that if the RTI data is normalized using a shell simulation with more sweeps, see Fig.\ \ref{fig:log_entropy}, then $\phik$ will shift to a yet lower value.  Thus the simulations suggest that a thermodynamic glass transition occurs at finite pressure.  This is the central physics result of this work.
         
    \subsection{Equilibration}
        \begin{figure}
            \centering
    		\includegraphics[width=0.45\textwidth]{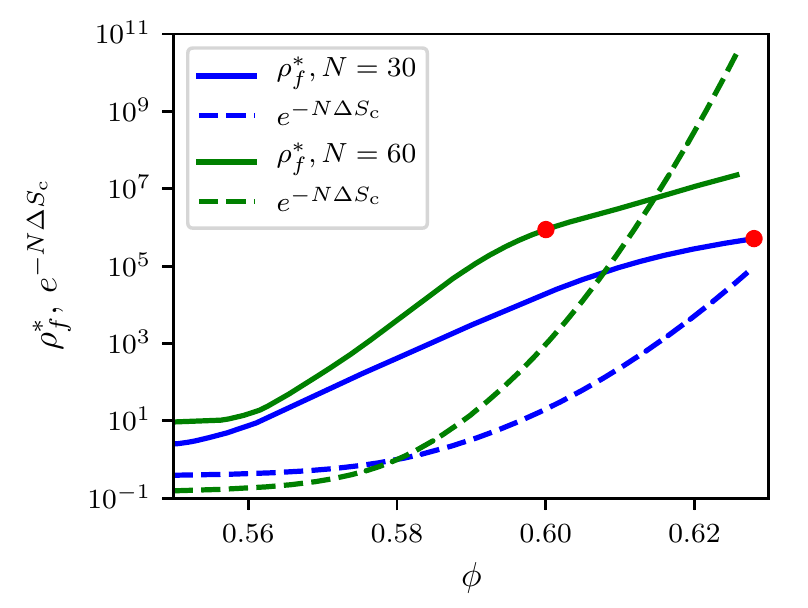}
            \caption{Plot of $\rhofs$ (solid) and $e^{-N\Delta\Sc}$ (dashed) vs $\p$ for $N=30$ (blue, lower) and $N=60$ (green, upper). Extrapolated values of $\Sc$ were used for $\phi > 0.59$. The red dots on the $\rhofs$ curves correspond to the point where the simulation falls out of equilibrium according to $\rhofs /R_\mathrm{tot} = 0.01$, where $R_{\mathrm{tot}}$ is the population summed over all simulations.
            }
            \label{fig:rhofs_Sc}
        \end{figure}   
        
        As discussed in Sec.\ \ref{sec:micro_PA_error}, one can estimate the systematic errors of a weighted average of many PA simulations using the quantity $\rhofs$ (see Eq.\ \ref{eq:rhofs}).  Following Ref.\ \cite{Callaham}, we set $\rhofs < 0.01 R_\mathrm{tot}$ as the threshold for equilibration of the weighted average of the simulations.  Here $R_\mathrm{tot}$ is the total population of the combined simulations. The solid lines in Fig.\ \ref{fig:rhofs_Sc} show $\rhofs$ as a function of packing fraction for $N=30$ (blue, lower) and $N=60$ (green, upper).  The equilibration threshold is $\rhofs= 6\times 10^5, \, 9.4 \times 10^5$ for $N=30, \, 60$ , respectively.  These thresholds are shown by red dots in the figure and suggest that the $N=30$ simulations are equilibrated over the entire range up to $\p=0.63$ while the $N=60$ results fall out of equilibrium at $\p=0.60$.   
        
        Figure \ref{fig:rhofs_Sc} shows a scatterplot of the joint distribution of the dimensionless pressure, $Z$, and total entropy, $\Stot$, at $\phi=0.60$  for the three system sizes.  Each point represents a single simulation included in the weighted average of observables and in the computation of $\rhofs$. The negative slope of the joint distribution shows that runs with higher values of $\Stot$ have lower values of $Z$, as expected.  In weighted averaging,  these high entropy runs are more heavily weighted so that the high entropy tail of the distribution must be well-sampled to accurately estimate $Z$, $\rhofs$ and $\Stot$.  Furthermore, for insufficient sampling, $\rhofs$ and $\Stot$ will tend to be underestimated and $Z$ overestimated.  It is clear that the high entropy tail is poorly sampled for the $N=60$ and $N=100$ runs but perhaps adequately sampled for $N=30$.  Thus, it seems likely that the range of equilibration suggested by the red dots in Fig.\ \ref{fig:rhofs_Sc} is overly optimistic due to insufficiently many runs used in estimating $\rhofs$. A more conservative approach would be to extrapolate $\rhofs$ using its nearly pure exponential behavior before the knee of the curve.  An exponential fit in the range before the knee of curve is insensitive to the precise fitting range. Setting the fitted function to the equilibration threshold yields more conservative estimates that equilibration is achieved for $\p \lesssim  0.618$ for $N=30$ and  $\p \lesssim 0.598$ for $N=60$.

        In Fig.\ \ref{fig:rhofs_Sc}, we also plot $e^{-N\Delta\Sc}$ vs $\phi$, where $\Delta\Sc(\phi) = \Sc(\phi) - \Sc(\phi^*)$.  In this expression we used the fitted value of $\Sc$ for $\phi > 0.59$, described in the previous section instead of the measured value since we believe the extrapolation of $\Sc$ is more accurate than the measured values deep in the glassy regime and leads to a more conservative criterion for equilibration. The initial packing fraction $\p^*=0.58$ is chosen to be close to the dynamic transition, $\p^* \approx \phid$, where each replica in the population is assumed to become trapped in a distinct glass state. For values of $\p$ such that $e^{-N\Delta\Sc}<0.01 R_\mathrm{tot}$ (shown in the plot by the height of the red dot), the combined simulations have sampled more than 100 equilibrium glass states. We see that this occurs over the whole range for which the $\rhofs < 0.01 R_\mathrm{tot}$ criteria is satisfied.

        \label{sec:results_equilib}
        \begin{figure}
            \centering
    		\includegraphics[width=0.45\textwidth]{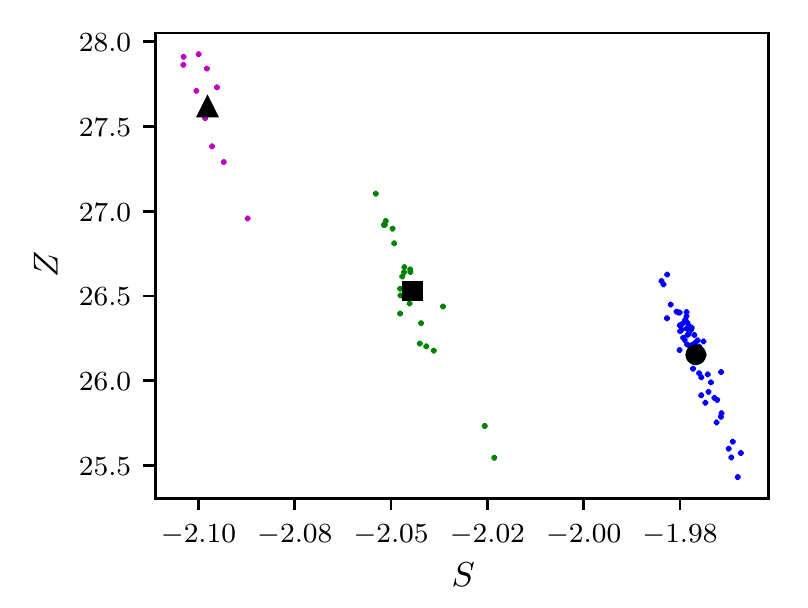}
    		\caption{ Scatter plot of dimensionless pressure $Z$ vs the total entropy per particle $\Stot$ at $\phi=0.60$ for $N=30$ (bottom, circle), $N=60$ (middle, square), and $N=100$ (top, triangle). Small symbols correspond to independent runs and large black symbols correspond to the weighted average of all runs.
    	    }
    		\label{fig:pressure_entropy}
    	\end{figure}

\section{Discussion}
\label{sec:disc}
    In this work we have introduced a new method to compute the configurational entropy of structural glasses and presented several new results relating to the glassy regime of binary hard sphere mixtures. Most importantly, we found new estimates of the Kauzmann density, $\phik$, and bounds to the random close packing density, $\prcp$. 
    The jamming density was obtained by fitting the pressure divergence and provides a lower bound on $\prcp$ that is slightly larger than previous estimates. The Kauzmann density was extrapolated from the configurational entropy measured in the equilibrium fluid regime beyond the dynamic glass transition, $\p>\phid$. Our results suggest that the Kauzmann transition occurs prior to random close packing, $\phik<\prcp$, so that a thermodynamic glass transition does indeed exist in this system.
    
    To obtain these results we introduced two new computational methods for calculating  the vibrational entropy of a configurational glass: the shell method, which is a variant of the Frenkel-Ladd method, and replica thermodynamic integration, which relies on integrating the entropy of individual replicas in population annealing starting from low density.  Replica thermodynamic integration produces a continuous curve of vibrational entropy but requires an alternative method to set an additive constant. 
    
    Population annealing has been shown to be capable of equilibrating configurational glasses beyond the dynamic glass transition \cite{Callaham}.  Here the dynamic transition is associated with the computational method, either Markov chain Monte Carlo or molecular dynamics, that drives equilibration in the algorithm at each packing fraction.  The limitation in probing equilibrium properties with population annealing is related to the precipitous decline in configurational entropy as the density is increased.  In population annealing, each member of the population is frozen in a glass state at $\phid$ so that upon entering the glass regime there are $R$ glass states, where $R$ is the population size. As the packing fraction increases, the number or distinct glass states in the population decreases exponentially as $R\, e^{N[\Sc(\p) - \Sc(\phid)]}$ and collapses into a single glass state when this number reaches one. The exponential dependence on $N$ explains why the method is restricted either to small systems or packing fractions only slightly above $\phid$.  On the other hand, population annealing and closely related techniques such as parallel tempering are the only methods known to us for going beyond the dynamic transition of a Monte Carlo or molecular dynamics scheme acting at a fixed packing fraction. It should be noted that the decrease in glass configurations with density imposes the same limitations on parallel tempering in going beyond the dynamic transition. However, in continuously polydisperse systems for which the swap algorithm is effective, population annealing (or parallel tempering) combined with the swap algorithm would permit direct measurements of equilibrium properties somewhat beyond the dynamic transition for the swap algorithm, which is already deep within the glassy regime.
    
\acknowledgments
    The work was supported in part by NSF Grant No. DMR-1507506.

\appendix
    \section{Entropy normalization} \label{app:entropy_norm}
        To find the normalization of the entropy per particle, we start with the thermodynamic definition,
        \begin{align}
            \frac{\partial S}{\partial V} = \frac{P}{T},
        \end{align}
        where the Boltzmann constant has been set to unity and $V$ is the volume per particle. This equation can be integrated in volume to get
        \begin{align}
            S(V) = \int_\Vig^{V} \frac{P}{T} dV' + S(\Vig),
        \end{align}
        where $\Vig$ is the ``ideal gas volume'' with the property that $\Vig \gg 1$. In this limit, we can make the approximation that the system begins as a binary mixture of two ideal gasses with entropy
        \begin{align}
            S(\Vig) = \log\left( \frac{\Vig}{\lambda_\mathrm{th}^3/2} \right) + \frac{5}{2},
        \end{align}
        where $\lambda_\mathrm{th}$ is the thermal deBroglie wavelength.
        By setting $\lambda_\mathrm{th}=1$, the resulting entropy at $V$ becomes
        \begin{align}
            S(V) = \lim_{\Vig \rightarrow \infty} \left[ \int_{\Vig}^{V} \frac{P}{T} dV' + \log( \Vig) \right] + \frac{5}{2} + \log(2).
        \end{align}
        We can rewrite this in terms of packing fraction by using the relationship between $\phi$ and $V$,
        \begin{align}
            \phi = \frac{4\pi}{3V} \frac{r_0^3 + r_1^3}{2},
        \end{align}
        where $r_0$ and $r_1$ are the radii of the small and large particles which, in our system, have values $r_0=1$ and $r_1=1.4$. Changing variables from $V$ to $\phi$ gives
        \begin{align}
            S(\phi) &= \lim_{\pig \rightarrow 0} -\int_{\pig}^\phi \frac{Z(\phi')}{\phi'} d\phi'  - \log(\pig) \\
            &+ \log\left( \frac{4\pi}{3}\frac{r_0^3 + r_1^3}{2} \right) + \frac{5}{2} + \log{2},\nonumber
        \end{align}
        where the $\log(\pig)$ term cancels with the logarithmically diverging pressure in the integral. By rearranging, we get a form without explicit divergences,
        \begin{align}
            \label{eq:app_entropy_1}
            S(\phi) &= \lim_{\pig \rightarrow 0} \int_{\pig}^\phi \frac{1-Z(\phi')}{\phi'} d\phi' - \log(\phi) \\
            &+ \log\left( \frac{4\pi}{3}\frac{r_0^3 + r_1^3}{2} \right) + \frac{5}{2} + \log{2}. \nonumber
        \end{align}
        Typically we do not integrate from $\phi=0$ and instead start at a non-zero initial packing fraction. Starting at $\p_0$, the total entropy in a population annealing simulation is given by
        \begin{align}
            S(\phi) &= S(\p_0) - \int_{\p_0}^\phi \frac{\tilde{Z}(\phi')}{\phi'} d\phi',
        \end{align}
        where $\tilde{Z}$ is the equilibrium pressure obtained during the population annealing simulation. The normalization entropy is given by Eq.\ \ref{eq:app_entropy_1},
        \begin{align}
            S(\p_0) = &\int_{0}^{\p_0} \frac{1-Z_\mathrm{BMCSL}(\phi')}{\phi'} d\phi' - \log(\p_0) \\
             &+ \log\left( \frac{4\pi}{3}\frac{r_0^3 + r_1^3}{2} \right) 
             + \frac{5}{2} + \log{2}, \nonumber            
        \end{align}
        where we have set $\pig$ to zero because the BMCSL equation of state, defined in Eq.\ \ref{eq:BMCSL}, can be explicitly integrated at $\p=0$.

\bibliographystyle{apsrev4-1}
\bibliography{refs,references}
\end{document}